\documentclass[12pt,a4paper,twocolumn]{article}

\usepackage[
  a4paper,
  top=1.8cm,
  bottom=2.0cm,
  left=1.7cm,
  right=1.7cm
]{geometry}

\usepackage{cuted}
\usepackage[T1]{fontenc}
\usepackage[utf8]{inputenc}
\usepackage[english]{babel}

\usepackage{lmodern}
\usepackage{microtype}

\usepackage{graphicx}
\usepackage[version=4]{mhchem}
\usepackage{amsmath,amssymb}
\usepackage{bm}
\usepackage{bbm}
\usepackage{csquotes}
\usepackage{enumitem}

\usepackage[
  backend=biber,
  style=phys,
  biblabel=brackets,
  citestyle=numeric-comp,
  sorting=none,
  minnames=4,
  maxnames=4
]{biblatex}
\usepackage{tikz}
\usetikzlibrary{arrows.meta,patterns.meta,calc}
\usepackage[
  colorlinks=true,
  linkcolor=blue,
  citecolor=blue,
  urlcolor=blue
]{hyperref}

\AtBeginDocument{%
}

\begin{document}

% ==================================================
% Full-width title and abstract
% ==================================================

\twocolumn[{%
\begin{minipage}{\textwidth}

  \centering

  {\LARGE\bfseries
  t-J model at 50  \par}

  \vspace{0.8em}

  {\large
  Józef Spałek
  \par}

  \vspace{0.6em}

  {\normalsize
  Institute of Theoretical Physics, Jagiellonian University,\\
  ul. Łojasiewicza 11, 30-348 Kraków, Poland\\
  \href{mailto:jozef.spalek@uj.edu.pl}
       {jozef.spalek@uj.edu.pl}
  \par}

  \vspace{0.6em}

  \vspace{1.0em}

  % Abstrakt o nieco mniejszej szerokości
  \begin{minipage}{0.92\textwidth}

    \small

    \noindent
    \textbf{Abstract.}
    I briefly overview the original formulation of the $t-J$ model starting from Hubbard model, as well as stress its unique
    features, i.e., advantages and shortcomings when discussing its physical properties. 
    Particular emphasis is put on a brief characterization of connection between 
    magnetism, unconventional superconductivity, and the Mott-Hubbard localization of the correlated
    carriers in a narrow band. In the second part I summarize the model generalization to the three--orbital situation. The main purpose is to list some basic features of the physics associated with $t-J$ model.
    \vspace{0.7em}

  \end{minipage}

  \vspace{1.2em}

\end{minipage}%
}]

% Od tego miejsca tekst jest automatycznie dwukolumnowy

\section{Introduction}

The present issue of Acta Physica Polonica A is devoted to the $40{\rm th}$ anniversary of the breakthrough discovery
of high temperature superconductivity in the cuprates. In this paper I would like to turn attention to the $50^{\rm th}$ anniversary of the $t-J$ model that turned to be intimately
connected to the subject, as well as  to discuss the author's contribution to its creation.

The following properties of the $t-J$ model of strongly correlated fermion systems determine its
fundamental role in condensed matter physics:
\begin{enumerate}[label=\roman* \!)]
    \item the model is directly relevant to high--temperature superconductivity by its 
    natural explanation of the real--space paired phase and its evolution from an antiferromagnetic insulating Mott state, as well as of competition with the emergence of charge--density--wave states;
    \item it makes possible the quantitative evolution of a strongly correlated metallic state
    to the Mott insulating phase in the half--filled--band limit. Hence, it combines the change of particle quantum statistics accompanying the transition from an itinerant to the localized state;
    \item it involves fermions with nontrivial (non--fermion) anticommutation relations and thus 
    allows to study non--Fermi (non--Landau) quantum (spin) liquids within a relevant microscopic model.
\end{enumerate}
Some of the aforementioned features are implicitly contained in more general models (e.g., the Hubbard model), but the starting point of the $t-J$ model makes it clear that new quantum states 
can appear already at the mean--field level. Some of these will be briefly elaborated here. Furthermore, the formal methodology
of the original derivation of the $t-J$ model can be directly applied to other models such as Anderson lattice or orbitally degenerate narrow--band models. 

The structure of this brief overview is as follows. In the Sects. 1.1. and 1.2, the Hubbard model and general forms of the $t-J$ model are discussed from the perspective of strong 
correlations. In Sect. 2, we discuss the main features from the perspective of related 
physics. In Sect. 3 we dwell upon generalizations of the methodology used in derivation and analysis of the original $t-J$ model. Sect. 5 contains brief a summary.

\subsection{Preliminaries: Interaction \\ either in  real or reciprocal space}
As a starting point of the theory of strongly correlated fermions, one can take the Hubbard's series of papers from the years 1963-1965 \cite{Hubbard1963,Hubbard1964canonical,Hubbard1965}. 
The point of departure here is  a single narrow band model formulated in the real--space (Wannier) representation $\{ | \mathbf{R}_{i \sigma}\rangle\} \equiv \{w_{i\sigma}(\pm) \equiv w_\sigma(\mathbf{r} - \mathbf{R}_i)\}$
of single--particle states, allowing us to discuss the correlations in the occupation--number representation (Fock space).
This formulation is contrasted with the standard approach in the (quasi)momentum representation 
$\{\mathbf{k}, \sigma \}\equiv\{ \Psi_{\mathbf{k} \sigma} (\mathbf{r}) \}$. The two representations are 
equivalent in the sense that the field operator $\hat{\Psi}_{\sigma} (\mathbf{r})$ can be defined
in either of them, i.e.,
\begin{align}
    \hat{\Psi}_\sigma(\mathbf{r}) \equiv \sum_i w_{i \sigma} (\mathbf{r}) \hat{a}_{i\sigma} \equiv \sum_{\mathbf{k}} \Psi_{\mathbf{k}\sigma} (\mathbf{r})\hat{a}_{\mathbf{k}\sigma},
\end{align}
since there is a unitary transformation between them, i.e.,
\begin{align}\label{eq12}
    \hat{\Psi}_\sigma (\mathbf{r}) = & \frac{1}{\sqrt{N}} \sum_l w_{i\sigma}(\mathbf{r})\sum_{\mathbf{k}} e^{i \mathbf{k}(\mathbf{r}-\mathbf{R}_i)} \hat{a}_{\mathbf{k}\sigma} \nonumber \\ \equiv &\sum_{\mathbf{k}} \Psi_{\mathbf{k}\sigma} \hat{a}_{\mathbf{k}\sigma},
\end{align}
as one takes that
\begin{align}
    \hat{a}_{\mathbf{k}\sigma} = \frac{1}{\sqrt{N}} \sum_i e^{i \mathbf{k}\mathbf{R}_i} \hat{a}_{i \sigma}.
\end{align}
Therefore, the expression for the two--particle (usually the Coulomb repulsive interaction) part
in the Fock space
\begin{align}\label{eq14}
    \hat{H}_2 &= \sum_{i j m n \sigma \sigma'} \int d^d r_1  d^d r_2 \hat{\Psi}_\sigma^\dagger (\mathbf{r}_1)
    \hat{\Psi}_\sigma^\dagger (\mathbf{r}_2) \nonumber \\ & V(\mathbf{r}_1 - \mathbf{r}_2)  \hat{\Psi}_{\sigma_2} (\mathbf{r}_2) \hat{\Psi}_{\sigma_1} (\mathbf{r}_1),
\end{align}
is equally valid in both the $\{ | \mathbf{k} \sigma \rangle\}$ and $\{ | i \sigma \rangle\} $ representations. However, by inserting explicit form (\ref{eq12}) to (\ref{eq14})
we obtain many terms that are very difficult to tackle on an equal footing. In effect,
one is forced to analyze simplified models in either representation,
which entails different approximation schemes from a physical side.

Starting from the \textbf{k}--representation one usually begins with the concept of an
electron gas or a simple band theory as a reference and then performs a Feynman--diagram expansion,
resulting in the Fermi liquid theory \cite{Abrikosov1963,PinesNozieres1966} or a Luttinger--liquid--type state in one dimension \cite{Luttinger1960}.
Additionally, within this framework, one also obtains Wigner--type crystallization in a model of a diluted Fermi gas \cite{Wigner1934}.

A quite different approach is employed when the whole analysis of extended (also periodic) systems is carried out in 
the Wannier (atomic) representation. In that situation we decompose the interaction 
into the intraatomic and interatomic parts. In effect, 
 (\ref{eq14}) takes the form of expansion
\begin{align}\label{eq15}
    \hat{H}_2 & = \frac{U}{2} \sum_{i  \sigma } \hat{n}_{i \sigma}\hat{n}_{i \overline{\sigma}} + \frac{1}{2} \sum_{i j}{'} K_{ij} \hat{n}_{i}\hat{n}_{j}\nonumber \\
    & - \frac{1}{2} \sum_{i j}{'} J_{ij}(\hat{\mathbf{S}}_{i}\cdot\hat{\mathbf{S}}_{j} + \frac{3}{4} \hat{n}_{i}\hat{n}_{j}) + \ldots,
\end{align} 
where $\hat{n}_{i \sigma}\!\equiv\!\hat{a}^\dagger_{i \sigma}\hat{a}_{i \sigma}$,  ~ $\hat{n}_{i}\!=\!\sum_\sigma \hat{a}^\dagger_{i \sigma}\hat{a}_{i \sigma}$, and $\hat{\mathbf{S}}_{i}\!\equiv\!(\hat{{S}}_{i}^+,\hat{{S}}_{i}^-,\hat{{S}}_{i}^z)$ $\equiv$ $\left(
\hat a^\dagger_{i\uparrow}\hat a_{i\downarrow},
\hat a^\dagger_{i\downarrow}\hat a_{i\uparrow},
\frac{1}{2}
(\hat n_{i\uparrow}-\hat n_{i\downarrow})
\right)$. The Hubbard intraatomic interaction part is characterized by $ U$, direct intersite Coulomb by $ K_{ij}$, and direct exchange 
interaction by $ J_{ij}$. For those interested in the full form of the Hamiltonian (with 
the inclusion of all two--site terms), as well as for explicit expressions for the coupling
constants in terms of  Wannier functions, see, e.g., Ref. \cite{Spalek2010}. In the present approach,
the interaction range is determined by the overlap of the Wannier functions. 
Explicitly, 
\begin{equation}\label{eq16}
    U \equiv \int d^3 r ~d^3r'  |w_{i \sigma} (\mathbf{r}) |^2 V(\mathbf{r} - \mathbf{r}') 
    |w_{i \overline{\sigma}} (\mathbf{r'})|^2,
\end{equation}
which in the case of the Coulomb interaction, when $w_{i \sigma} (\mathbf{r}) = w_{i} (\mathbf{r}) \chi_\sigma $, takes the form
\begin{equation}\label{eq17}
    U \equiv \int d^3 r ~d^3r' e^2  \frac{|w_{i}(\mathbf{r})|^2  |w_{i} (\mathbf{r'})|^2}{|\mathbf{r} - \mathbf{r}'|}.
\end{equation}
Likewise, 
\begin{equation}
    K_{ij} \equiv \int d^3 r~ d^3 r' e^2  \frac{|w_{i }(\mathbf{r})|^2  |w_{j} (\mathbf{r'})|^2}{|\mathbf{r} - \mathbf{r}'|}.
\end{equation}
In both above expressions above, the leading terms of interaction have the density--density form with the effective charge density $e\cdot |w_i(\mathbf{r})|^2$.

If the corresponding densities are largely concentrated near
their parent atomic site positions $\mathbf{R}_i$ and $\mathbf{R}_j$, we can truncate the interaction to
either to the intraatomic part alone (the Hubbard model) or extend it to the nearest neighbors $(i,j)$. These are 
the assumptions made in almost all theoretical papers on correlated systems, which on one hand lead to results that are often not easily comparable with those obtained when starting from  the concept of a fermionic gas. However, the alternative approach leads to a series of basic results, some of which are discussed later.

In a similar manner, one can define the analog of the Hubbard model, in $\mathbf{k}$ space
\begin{equation}\label{eq191}
    \hat{\mathcal{H}} = \sum_{\mathbf{k}\sigma} \epsilon_{\mathbf{k}} \hat{n}_{\mathbf{k}\sigma} + \frac{U}{2} \sum_{\mathbf{k}\sigma } \hat{n}_{\mathbf{k}\sigma}\hat{n}_{\mathbf{k}\overline{\sigma}} + \ldots,
\end{equation}
where $\epsilon_{\mathbf{k}}$ is a single--particle energy and the dots at the end
suggest that this model can be extended in a similar to the previous case. The interesting aspect of this approach is that (\ref{eq191}), in its simplest form, can be diagonalized exactly. On the basis of the exact solution, one can build up Fermi--liquid--type corrections
of the Landau--Fermi--liquid type, but that aspect will not be touched upon here.

Let us mention only that the original Hubbard--model validity is limited to the situation
with small overlap of the neighboring Wannier states $w_i(\mathbf{r})$ and $w_{j(i)}(\mathbf{r})$, whereas the applicability of (\ref{eq191}) is well suited to situation involving almost singular elastic scattering amplitude $U$ of the states with  $k=k'$. So, the situations reflect
different limits. Nevertheless, it is intriguing to ask if there are any universal physical results coming from these two complementary ($\mathbf{k}$-- and real-- space) models. This question will be the subject of our group analysis in the 
nearest future.

\subsection{Anderson kinetic exchange for Mott insulators (1959)}
The direct exchange interactions given in (\ref{eq15}) were introduced to physics by Fock \cite{Fock1932,FockSelectedWorks2004}.
Note that here, the spin operator $\hat{\mathbf{S}}_i$ has been expressed by fermion operators.
This form differs from the original analyses by either Heisenberg \cite{Heisenberg1928} or  Dirac \cite{Dirac2010canonical}, where the spin-$1/2$ operators are just atomic spins $\hat{\mathbf{S}}_i= \frac{1}{2} \hat{\boldsymbol{\tau}}_i$, and $\boldsymbol{\tau}_i$ are Pauli matrices, representing the spin
of electron localized on atom located at position $i\equiv \mathbf{R}_i$. The value of $J_{ij}$
is then such that,  in a many--particle system it usually favors a ferromagnetic ordering.
The main problem then in the physics of transition--metal oxides was then that even though they possess
localized $3d$ electrons on the atoms, the exchange interaction was -- as a rule -- antiferromagnetic and quite strong.
 This problem has been resolved by Anderson \cite{Anderson1959canonical,Anderson1963} who explained this fact within the real--space representation of the single--particle states. Although it was 
 not explicitly said in the original paper \cite{Anderson1959canonical}, Anderson implicitly introduced the Hubbard model, as a starting point of his considerations, i.e.,
 \begin{equation}\label{eq19}
     \hat{\mathcal{H}} = \sum_{ i j \sigma}{'} t_{ij} \hat{a}_{i \sigma}^\dagger \hat{a}_{j \sigma} + U \sum_i \hat{n}_{i \uparrow}^\dagger \hat{n}_{i \downarrow},
 \end{equation}
 where $t_{ij} < 0 $ is the hopping integral between sites $i$  and $j$, and the primed summation
 means $i \neq j$. Starting from this point, he calculated the form of the effective spin Hamiltonian in the second order in $t_{ij}$ of perturbation theory, i.e.,
 \begin{equation}\label{eq110}
     + \frac{1}{2} \sum_{ij}{'} \frac{2 t_{ij}^2}{U} (\hat{\mathbf{S}}_i \cdot \hat{\mathbf{S}}_j
 - \frac{1}{4}).
 \end{equation}
This interaction is called the \textit{kinetic exchange}, since the exchange integral $J_{ij}$ is expressed via the kinetic motion of electrons ($\sim t_{ij}$) within a lattice of essentially localized spins.

In reality, the term $\sim t_{ij}^2$, represents only the virtual  electron hopping $i\leftrightarrow j$, i.e., forth and back between the sites $i$ and $j$. The question then is: What is the influence of direct hopping term $i\to j$ onto the system's energy? In other words, how does the Eq.
(\ref{eq19}) reduces to (\ref{eq110}) in  the limit of the Mott insulating state? The subsequent 
treatment of (\ref{eq19}) by Hubbard \cite{Hubbard1963,Hubbard1964canonical,Hubbard1965} tackled this problem, albeit only in the paramagnetic state only. Also, within his 
approach, Hubbard could not obtain an explicit transformation of (\ref{eq19}) into (\ref{eq110}). Additionally, the fundamental question arises: What form, if any, might this kinetic exchange  interaction can take in a correlated metallic state? This problem was solved by the present author and published in cooperation with 
his colleagues in a series of papers \cite{SpalekOles1976,spalek1977magnetic,Chao1977canonical,SpalekChaoOles1978,Spalek1981Habilitation}. Next, as we discuss this before turning to its concrete subsequent applications.
%break
\section{The $t-J$ model as such: Single--orbital band case}
The essence of strong--correlation physics analyzed via Hubbard or other related 
models, relies on the assumption that the intraatomic interaction $\sim \! U$ defines the largest 
energy scale in the system, i.e. $U\gg |t_{ij}|, K_{ij}, J_{ij}$, etc. This presumption automatically
means that ordinary perturbation expansion, in which we start from single--particle states 
(as defined by, e.g., Slater determinant composed of individual Wannier functions) is not adequate, if applicable at all. 
\begin{figure*}[htb]
\centering
\includegraphics[width=0.9\textwidth]{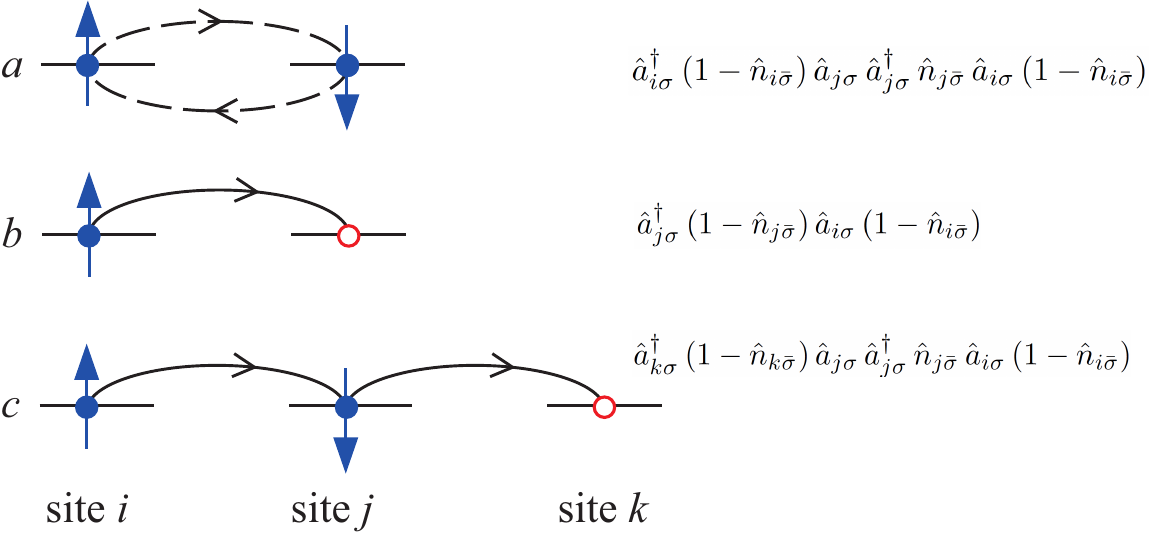}
\caption{Various neighboring hopping processes: a virtual hopping, b-real hopping, and c-three-site hopping in the second order. In the limit of Mott insulator only %the first term (a)
survives and leads to the pure kinetic-exchange interaction form.}
\label{Fig2}
\end{figure*}
To overcome this obstacle, one resorts to dividing the Fock space into the following subspaces: the lowest subspace with empty
or singly occupied sites, and that with at least one doubly occupied site, etc. In effect, we have a ladder of subspaces characterized by $l = 0,1,2,\ldots$ etc. doubly occupied sites, with the gaps 
in between of them of magnitude $U$ in the limit of atomic limit. Accordingly, one has to divide the hopping into two parts: Hops from singly occupied to empty sites and those between singly occupied sites with opposite spins. These two types of hopping (in the first order,
$\sim t_{ij}$), are shown schematically in Fig. 1. From the schematic representation one can clearly see that the process (a) is taking place to a higher-energy state, so it can effectively influence the system dynamics only via virtual hops forth and back, 
which provide a contribution $\sim t_{ij}^2$. Such a replacement of highly excited 
states via virtual hopping represents a typical way of obtaining the effective (low--energy) 
model once the whole operation is carried out in an invariant (operator) form, i.e., 
must be performed via the canonical perturbation expansion in the proper (operator) form.

Formally (cf. Fig. 2), the whole procedure is performed in two steps. First, one makes a decomposition 
of the occupation--number and fermion operators as
\begin{align}
    \hat{n}_{i \sigma} \equiv \hat{n}_{i \sigma} (1 - \hat{n}_{i \overline{\sigma}} + \hat{n}_{i {\sigma}}) = \hat{n}_{i \sigma} (1 - \hat{n}_{i \overline{\sigma}}) + \hat{n}_{i \sigma} \hat{n}_{i \overline{\sigma}},
\end{align}
and
\begin{align}
    \hat{a}_{i \sigma} &\equiv \hat{a}_{i \sigma} (1 - \hat{n}_{i \overline{\sigma}} + \hat{n}_{i {\sigma}}) \equiv \nonumber \\ &\hat{a}_{i \sigma} (1 - \hat{n}_{i \overline{\sigma}}) + \hat{a}_{i \sigma} \hat{n}_{i \overline{\sigma}} \equiv \hat{b}_{i \sigma} + \hat{d}_{i {\sigma}}.
\end{align}

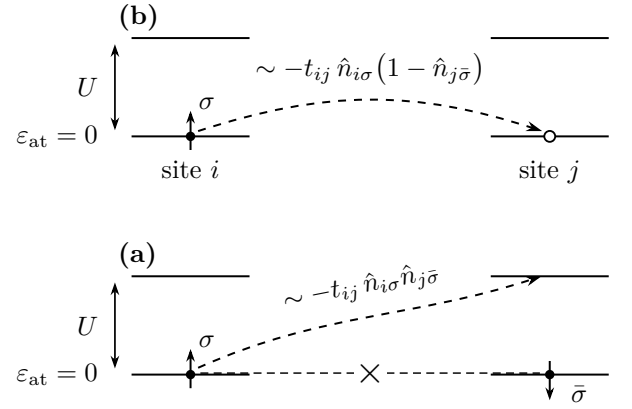
\begin{figure}[t]
\centering
\begin{tikzpicture}[
    x=1cm,
    y=1cm,
    level/.style={line width=0.75pt},
    hop/.style={
        dashed,
        line width=0.75pt,
        -{Stealth[length=2.1mm,width=1.35mm]}
    },
    blocked/.style={densely dashed,line width=0.55pt},
    spin/.style={
        line width=0.75pt,
        -{Stealth[length=1.7mm,width=1.15mm]}
    },
    energy/.style={
        line width=0.65pt,
        <->,
        >={Stealth[length=1.7mm,width=1.15mm]}
    },
    electron/.style={circle,fill,inner sep=1.25pt},
    vacancy/.style={circle,draw,line width=0.7pt,fill=white,inner sep=1.45pt},
    every node/.style={font=\footnotesize}
]

% Geometry
\def\xleft{0}
\def\xright{4.75}
\def\halflevel{0.78}
\def\Ehigh{1.30}

% =============================================================
% (a) singly occupied -> empty, within the low-energy subspace
% =============================================================
\begin{scope}[yshift=3.15cm]

    \node[font=\bfseries\footnotesize,anchor=west]
        at (-1.10,1.58) {(b)};

    % Atomic levels on sites i and j
    \foreach \x in {\xleft,\xright}{
        \draw[level] (\x-\halflevel,0) -- (\x+\halflevel,0);
        \draw[level] (\x-\halflevel,\Ehigh) -- (\x+\halflevel,\Ehigh);
    }

    % Atomic excitation energy
    \draw[energy]
        (-1.00,0.08) -- node[left=2pt] {$U$} (-1.00,\Ehigh-0.08);
    \node[anchor=east] at (-1.08,0) {$\varepsilon_{\rm at}=0$};

    % Initial electron on i
    \node[electron] (ai) at (\xleft,0) {};
    \draw[spin] (\xleft,-0.18) -- (\xleft,0.34);
    \node[above right=-1pt] at (\xleft,0.25) {$\sigma$};

    % Empty target site j
    \node[vacancy] (aj) at (\xright,0) {};

    % Allowed first-order hopping
    \draw[hop]
        ($(ai)+(0.10,0.07)$)
        to[out=18,in=162]
        node[midway,above=4pt,fill=white,inner sep=1.2pt]
        {$\displaystyle \sim -t_{ij}\,\hat n_{i\sigma}
          \bigl(1-\hat n_{j\bar\sigma}\bigr)$}
        ($(aj)+(-0.10,0.07)$);

    \node[align=center,below=5pt] at (\xleft,0)
        {site $i$};
    \node[align=center,below=5pt] at (\xright,0)
        {site $j$};

\end{scope}

% =============================================================
% (b) two singly occupied sites -> virtual doublon at energy U
% =============================================================
\begin{scope}

    \node[font=\bfseries\footnotesize,anchor=west]
        at (-1.10,1.58) {(a)};

    % Atomic levels on sites i and j
    \foreach \x in {\xleft,\xright}{
        \draw[level] (\x-\halflevel,0) -- (\x+\halflevel,0);
        \draw[level] (\x-\halflevel,\Ehigh) -- (\x+\halflevel,\Ehigh);
    }

    % Atomic excitation energy
    \draw[energy]
        (-1.00,0.08) -- node[left=2pt] {$U$} (-1.00,\Ehigh-0.08);
    \node[anchor=east] at (-1.08,0) {$\varepsilon_{\rm at}=0$};

    % Initial state: i sigma, j bar-sigma
    \node[electron] (bi) at (\xleft,0) {};
    \draw[spin] (\xleft,-0.18) -- (\xleft,0.34);
    \node[above right=-1pt] at (\xleft,0.25) {$\sigma$};

    \node[electron] (bj) at (\xright,0) {};
    \draw[spin] (\xright,0.18) -- (\xright,-0.34);
    \node[anchor=west] at (\xright+0.14,-0.25) {$\bar\sigma$};

    % A direct hop cannot remain in the low-energy subspace.
    \draw[blocked]
        ($(bi)+(0.13,0.02)$) -- ($(bj)+(-0.13,0.02)$);
    \node[font=\large,fill=white,inner sep=0.5pt]
        at ({0.5*(\xleft+\xright)},0.02) {$\times$};

    % Hopping into the high-energy, doubly occupied sector
    \coordinate (doublon) at (\xright,\Ehigh);
    \draw[hop]
        ($(bi)+(0.10,0.09)$)
        to[out=24,in=198]
        node[midway,above=5pt,sloped,fill=white,inner sep=1.2pt]
        {$\displaystyle \sim -t_{ij}\,\hat n_{i\sigma}
          \hat n_{j\bar\sigma}$}
        ($(doublon)+(-0.11,0.00)$);

    % Virtual doublon: opposite spins occupying site j
    %\node[electron] at ($(doublon)+(-0.07,0)$) {};
    %\node[electron] at ($(doublon)+( 0.07,0)$) {};
    %\draw[spin]
    %    ($(doublon)+(-0.07,-0.16)$) -- ($(doublon)+(-0.07,0.30)$);
    %\draw[spin]
    %    ($(doublon)+( 0.07,0.16)$) -- ($(doublon)+( 0.07,-0.30)$);
    %\node[anchor=east] at ($(doublon)+(-0.16,0.28)$) {$\sigma$};
    %\node[anchor=west] at ($(doublon)+(0.16,-0.24)$) {$\bar\sigma$};
    %\node[align=center,above=8pt] at (\xright,\Ehigh)
    %    {virtual doublon};

    %\node[align=center,below=5pt] at (\xleft,0)
    %    {site $i$\\singly occupied};
    %\node[align=center,below=5pt] at (\xright,0)
    %    {site $j$\\singly occupied};

\end{scope}
\end{tikzpicture}
\caption{
Hopping between neighboring sites $(i,j)$ in the first order:
singly occupied to empty (a) and between singly occupied sites
$(i\sigma,j\bar{\sigma})$; the latter to the higher-energy state,
as $|t_{ij}|\ll U$ is assumed. The Figure shows explicitly the matrix elements
not shown in Fig. 1.
}
\label{fig:hopping-processes}
\end{figure}
Note that the operators represent two orthogonal subspaces as for every site
\begin{align}
    &\hat{b}_{i \sigma} \cdot \hat{d}_{i \sigma} \equiv 0, \nonumber \\ & \hat{\nu}_{i \sigma} \hat{d}^\dagger_{i \sigma} \hat{d}_{i \sigma} \equiv 0,
\end{align}
so the Fock subspaces with different numbers of doubly occupancies compose a proper Fock subspace each.
Nonetheless, the hopping between them mixes the subspaces, i.e.,
\begin{align}
    t_{ij} \hat{a}_{i \sigma}^\dagger \hat{a}_{j \sigma} &\equiv  t_{ij} \nonumber 
    \left[ \hat{a}_{i \sigma}^\dagger (1 - \hat{n}_{i \overline{\sigma}})\hat{a}_{j \sigma} (1 - \hat{n}_{j \overline{\sigma}}) \right. + \\ 
    &\left. \hat{a}_{i \sigma}^\dagger  \hat{n}_{i \overline{\sigma}} \hat{a}_{j \sigma} \hat{n}_{j \overline{\sigma}}   + \hat{a}_{i \sigma}^\dagger (1 - \hat{n}_{i \overline{\sigma}})\hat{a}_{j \sigma} \hat{n}_{j \overline{\sigma}} \right. \nonumber \\ &\left. + \hat{a}_{i \sigma}^\dagger \hat{n}_{i \overline{\sigma}} \hat{a}_{j \sigma} (1 - \hat{n}_{j \overline{\sigma}}) \right].
\end{align}
The first two processes do not mix the subspaces, whereas the remaining two do. So, to 
recover the orthogonality of the subspaces one has to remove the latter via the corresponding canonical transformation in an invariant (operator) form. The solution of this particular problem presented itself a remarkable obstacle to the author. This was particularly so because the part regarded as a "perturbation", i.e., the part
\begin{equation}
    \mathcal{\hat{H}}_1 \equiv \sum_{i j \sigma}{'} t_{ij} (\hat{b}_{i \sigma}  \hat{d}_{i \sigma} + \hat{d}_{i \sigma}^\dagger  \hat{b}_{i \sigma}),
\end{equation}
contains the term $\sim t_{ij}$, which appears also in "unperturbed" part
\begin{align} \label{eq117}
    \mathcal{\hat{H}}_0 \equiv \sum_{i j \sigma}{'} t_{ij} (\hat{b}_{i \sigma}^\dagger \hat{b}_{j \sigma} + \hat{d}_{i \sigma}^\dagger  \hat{d}_{j \sigma}) + 
    \frac{U}{2} \sum_\sigma \hat{d}^\dagger_{i \sigma} \hat{d}_{i \sigma}.
\end{align}
Note that here $\hat{d}^\dagger_{i \sigma} \hat{d}_{i \sigma} \equiv \hat{d}^\dagger_{i \overline{\sigma}} \hat{d}_{i \overline{\sigma}}  $, as we can change in the summation index $\sigma \to \overline{\sigma}$. In effect, 
$\langle \hat{d}^\dagger_{i \sigma} \hat{d}_{i \sigma} \rangle \equiv d_i^2$.

The further procedure of obtaining the effective Hamiltonian in the second order 
($t-J$ model) has been presented and overviewed in detail elsewhere  \cite{Anderson1959canonical,Anderson1963,SpalekOles1976,spalek1977magnetic,Chao1977canonical,SpalekChaoOles1978,Spalek1981Habilitation,Spalek2022}. Below we provide the resultant effective Hamiltonian and elaborate on its most important features. The full form 
of the Hamiltonian contains both two-- and three-- site terms (see also Appendix A), which can  be brought
to a simpler form, see below.

\subsection{The $t-J$ model: Original form (1976, see Fig. 3)}
We write down the effective Hamiltonian in the somewhat more involved form, containing all two--site 
interaction terms, i.e., those of the extended Hubbard model. 
Namely, for the lowest Hubbard subband (applicable for band filling 
$n \equiv \langle \hat{n}_i \rangle \le 1$) it has the form
\begin{align}\label{eq118}
    \hat{\widetilde{\mathcal{H}}} \equiv \sum_{i j \sigma}\!{'} t_{ij} \hat{b}_{i \sigma}^\dagger \hat{b}_{j \sigma} + \frac{1}{2} \sum_{ij \sigma \sigma}\!{'} (K_{ij} - \frac{1}{2} J_{ij}) \hat{\nu}_{i \sigma}\hat{\nu}_{j \sigma} \nonumber \\+ \sum_{ij}\!{'} J_{ij} \hat{\mathbf{S}}_i \hat{\mathbf{S}}_j + \rm{(three-site ~ terms)},
\end{align}
with effective kinetic exchange constant
\begin{equation}
    J_{ij} \equiv \frac{2 (t_i + V_{ij})^2}{U - K_{ij} + \frac{1}{2}J_{ij}^H} - J_{ij}^H.
\end{equation}
As before, $K_{ij}$ is the constant of the direct intersite Coulomb interaction, whereas
$J_{ij}^H$ is the direct (Heisenberg--Dirac) exchange constant, and $V_{ij}$ represents the so--called correlated hopping magnitude. I am citing this form from the less known publications \cite{Spalek2022,Spalek1978a} to show explicitly the condition under which the kinetic exchange 
interaction becomes dominant over the direct Heisenberg--Dirac interaction $\sim J^H$. 
Obviously, a realistic estimate of its value would require an explicit evaluation of 
the microscopic parameters $t_{ij}$, $U$, $K_{ij}$, $J_{ij}^H$, and $V_{ij}$, based on a concomitant calculation of the single--particle orbitals $\{ w_i (\mathbf{r})\}$ \cite{Spalek2020sorbitals}. Having in mind that model (\ref{eq15}) and all that follows on its basis, we consider here only a parametrized version of our microscopic model.
\begin{figure*}[!htbp]
\centering
\includegraphics[width=0.88\textwidth]{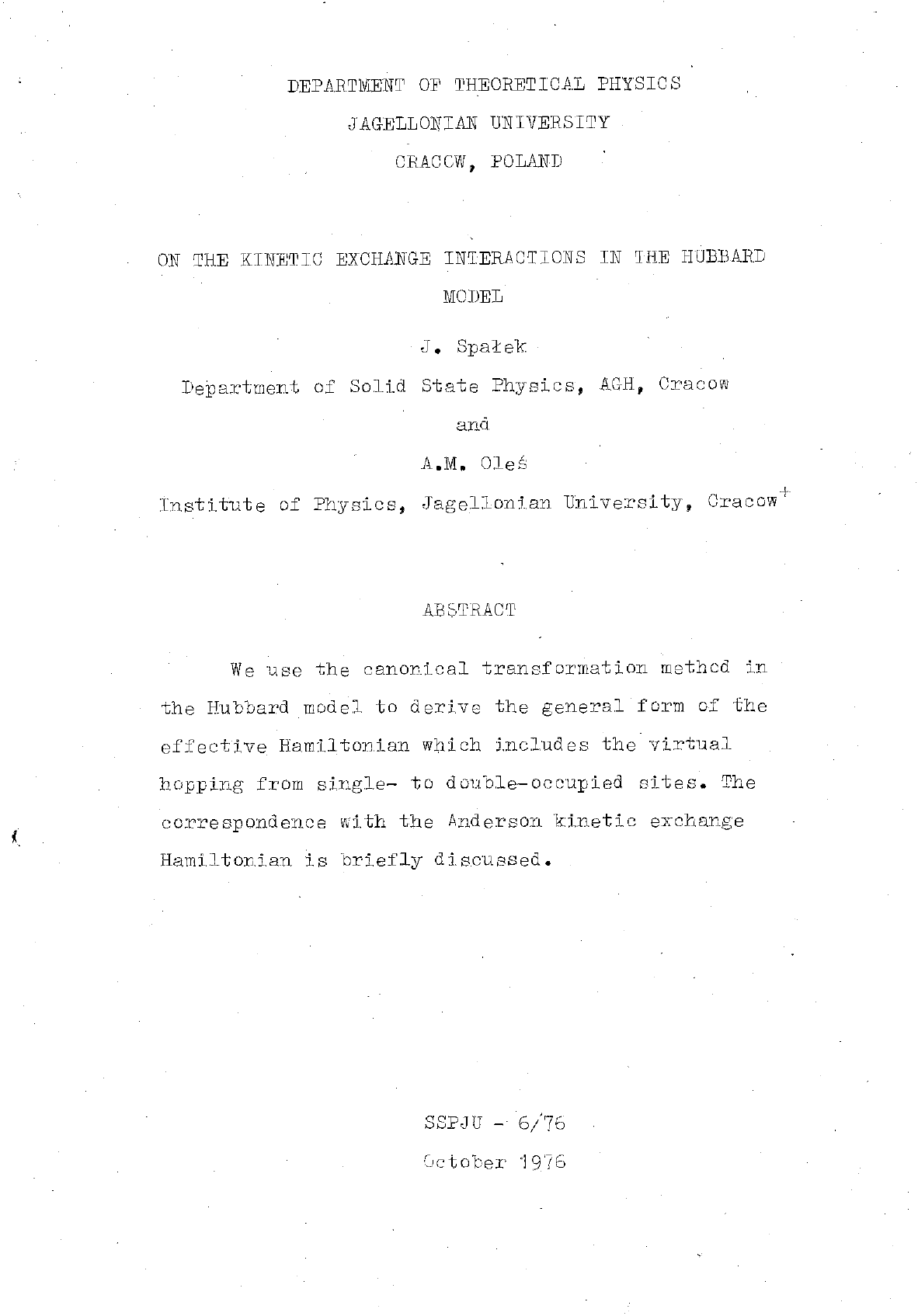}
\caption{The first page of the unpublished report \cite{SpalekOles1976} on the derivation of the $t-J$ model (see doi:10.26106/4s8w-hv63).}
\label{fig:12}
\end{figure*}
\subsection{$t-J$ model: Characteristics }
Physical comments are in place here.
First, as just said the Hubbard model (or its extended versions) is not fully microscopic, since the microscopic parameters are not explicitly evaluated. Its usefulness is nonetheless connected 
with the fact that, with the help of a few parameters ($t/U, J/U$) one can characterize quite a number of physical properties:
Mott--Hubbard (metal--insulator) transitions, high--temperature superconductivity, quantum spin--liquid for almost localized Fermi or non--Fermi liquids, and spin--dependent masses \cite{Spalek2026SCQMoverview}, all under the supposition that the parameters do not change with the
concentration of electrons (band filling ) $n$, or with temperature. One has to keep that 
in mind when applying this and related parametrized microscopic models to real quantum materials.

 Additional points, usually ignored, should also be mentioned. First, the higher--order 
corrections to the model lead to a further renormalization of the effective hopping 
integral ($t_{ij}$).  Second, in the fourth order of the canonical transformation,
biquadratic exchange interaction $\sim (\mathbf{S}_i \cdot \mathbf{S}_j) (\mathbf{S}_k \cdot \mathbf{S}_l)$ appear in a natural manner. 
Also a nonstandard form of $t-J$ model is obtained when electron--lattice interactions (e.g., the latter in the Rashba form) are included. Those additions are necessary to discussing 
strongly correlated chalcogenides such as $W Se_2$, $IrSe_2$, etc. We will not dwell here upon those recent developments.

The next remark concerns the real (dimensionless) coupling parameters. In the Hubbard model
it is $|t_{ij}|/U$ (in extended version $|t_{ij}|/(U-K)$). Earlier, it was thought,
also by the author, that a proper such parameter is $W/(U - K_{ij})$, where W is the 
bare bandwidth, $W \equiv |\sum_{j(i)} t_{ij}|$.
However, as the summations in the effective model (\ref{eq118}) have the same form, one sees 
that effectively the condition $|t_{ij}|/(U-K)\ll 1$ guarantees the fast convergence of the expansion (canonical perturbation) in $t_{ij}/(U - K_{ij})$. In effect, we can presume that the expansion is roughly in $\frac{1}{2 z} (W/U)$, where $z$ is the number of nearest neighbors. Hence, the model
may be regarded as applicable even when $W/(U-K)\sim 1$.

Finally, from what has been said above one can infer that in $t-J$ model it may happen 
that the hopping and exchange parts may become of comparable magnitude, 
particularly when the spin--singlet correlations $\langle \hat{\mathbf{S}}_i\hat{\mathbf{S}}_j - \frac{1}{4} \hat{\nu}_i \hat{\nu}_j \rangle \sim 1$, 
i.e., are strong. This is because the hopping part, expressed in projection operators 
gives a contribution $\langle \hat{b}^\dagger_{i \sigma} \hat{b}_{j \sigma} \rangle \sim (1 - n) \langle \hat{a}^\dagger_{i \sigma} \hat{a}_{j \sigma} \rangle$, where (1-n) is the so--called hole filling (deviation from the half--filling).
Under these circumstances the magnitude of the hopping is $0.1 ~t_{ij}$ for $n = 0.9 < J_{ij}$!.

Explicitly, for the cuprate high--temperature systems, taking the leading parameters $|t_{ij}| = 0.35 ~ eV$, along with $U = 10 ~ eV$, and $K_{\langle ij\rangle} = (1/3) U \simeq 3 ~ eV$, one obtains $J_{\langle ij\rangle} \equiv 4 t^2_{\langle ij \rangle} / (U - K_{\langle ij \rangle}) \sim 0.15 ~ eV$, 
whereas $(1 - n) t_{ij} \approx 0.04 ~ eV \sim \frac{1}{3} J_{\langle ij \rangle}$. Thus, 
we have that $|t_{\langle ij \rangle}| / (U - K) \simeq 1/4$ and the model is  valid for those
strongly correlated particles, i.e., for a liquid of hopping spins and fermion holes, with magnitude of spin hopping (coming from the three--site term) is $\sim 2 t^2_{\langle ij \rangle}/(U - K_{\langle ij \rangle}) (1-n) \sim 0.02 ~ eV$. This mixed spin--hole liquid freezes into a spin--fluctuating liquid or frozen--spin antiferromagnetic state, depending on the system spatial dimension $d \leq 2$.

One formal, but important, feature of the $t-J$ model must mentioned. Namely, the Hamiltonian (\ref{eq118}) should still be projected globally onto the subspace of states excluding totally, not only locally, the 
double occupancies. This is carried out by introducing the global projector $P_1$ and then instead of (\ref{eq118}) we should take the form $P_1 \widetilde{\mathcal{H}}P_1 \equiv \widetilde{\mathcal{H}}$ in what follows. We have
so far disregarded this feature and the reason for that will become clear when we discuss an effective $t-J-U$ model later \cite{Dey2026a}. Equivalently, one can take the globally projected wave function 
in the form $P_1 |{\Psi_0}\rangle$ and this procedure will do the job, particularly in variational calculations.

\subsection{The $t-J$ model: From correlated spins to local pairs}
The renewed interest in the $t-J$ model in the 1980s was largely associated with a qualitative suggestion of Anderson \cite{Anderson1987} that the antiferromagnetic kinetic exchange can be directly related to the local (real--space)  spin--singlet pairing. The present author was immediately inspired by that concept
and have invoked \cite{Spalek1988} the exact formal expression of that Anderson's statement. Explicitly,
one introduces the following local projected pair operators
\begin{align}
\left\{
\begin{aligned}
    \hat{B}_{ij} &\equiv \frac{1}{\sqrt{2}} (\hat{b}_{i\uparrow}\hat{b}_{j\downarrow} - \hat{b}_{i\downarrow}\hat{b}_{j\uparrow})  \\
    \hat{B}_{ij}^\dagger &\equiv (\hat{B}_{ij})^\dagger.
    \end{aligned}
\right.
\end{align}
Then the $t-J$ model in its standard form, (e.g., for $K_{ij} = J_{ij}^H = 0$)  takes the compact form even when we include the three--site terms not explicitly written in (\ref{eq118}), namely
\begin{align}
    \hat{\widetilde{\mathcal{H}}} &\equiv \sum_{ij\sigma}\!{'}  t_{ij} \hat{b}_{i\sigma}^\dagger \hat{b}_{j\sigma} \nonumber \\
    & - \sum_{ijk}\!{''}  \frac{2 t_{ij} t_{jk}}{U} \hat{B}_{ij}^\dagger\hat{B}_{kj} .
\end{align}

In this form, the pairing part lowers the energy when the local spin--singlet pairs are formed.
However, this also means automatically that any spin--singlet paired state 
will compete with spin density state appearance, as can be explicitly seen from the equivalence
\begin{align}
    P_1 \hat{B}_{ij}^\dagger  \hat{B}_{ij} P_1 \equiv -P_1 (\hat{\mathbf{S}}_i \cdot \hat{\mathbf{S}}_j - \frac{1}{4} \hat{n}_{i}\hat{n}_{j}) P_1.
\end{align}
Parenthetically, the r.h.s. part
of this formula can be regarded as a projection of any fermion--pair state onto 
its singlet part. Also, the charge--density correlations may arise from the $P_1 \hat{n}_i\hat{n}_j P_1$ term, particularly if the direct intersite Coulomb 
interaction $\sim K_{ij}$ is included in the extended form of the model. 

The full form of the pairing $t-J$ Hamiltonian with three-site terms included is given by \cite{Spalek1988}
\begin{align}\label{eq222}
    \hat{\widetilde{\mathcal{H}}} &= P_1 \biggl\{ \sum_{ij\sigma}\!{'} t_{ij} \hat{b}_{i\sigma}^\dagger \hat{b}_{j\sigma} \nonumber \\
    & - \sum_{ijk}\!{''}  \frac{2 t_{ij} t_{jk}}{U} \hat{B}_{ij}^\dagger\hat{B}_{kj} \biggl\}P_1.
\end{align}
The double primed summation means that $i\neq j\neq k$.
Physically, one can say that in this form the correlated single--particle hopping is entangled  with the dynamic of singlet pairs (pair hopping via $k \neq j$), as discussed elsewhere \cite{SpalekGoc2012}. Note that global projection operator $P_1$ is, in general, necessary if we are to talk sensibly about 
physical properties in the strong correlation limit, i.e., $|t_{ij}| \ll U$, when the double occupancies are excluded
at \textit{all} sites.
The importance of this last remark may be illustrated  by the example of taking the Mott--insulator limit, when not only $\langle \hat{n}_i\rangle = \langle \hat{\nu}_i \rangle = 1$, 
but also $\hat{n}_{i \uparrow} + \hat{n}_{i \downarrow} =  \hat{\nu}_{i \uparrow} + \hat{\nu}_{i \downarrow} = 1$ in the operator form, implying that the number of particles is conserved at each site. Under this circumstance the hopping part vanishes and the $t-J$ model reduces exactly to the Anderson kinetic exchange counterpart (\ref{eq110}), as it should be.

%break zrobic cytowania 
\subsection{From $t-J$ to $t-J-U$ model: A formal extension}

As said above, the $t-J$ model has specific novel features when
compared to the original (or extended) Hubbard model. The attractiveness of the former stems from the property that the spin--and/or pairing--orderings to show up 
explicitly on the mean--field level. This is to say that a paired phase does not appear in the mean--field (Hartree--Fock) approximation of the Hubbard model. The mean--field (SGA \cite{JedrakSpalek2010,JedrakSpalek2011})
solution provides one and this has been discussed in very many papers (e.g., see \cite{JedrakSpalek2010,JedrakSpalek2011} for a solution with the statistically consistent Gutzwiller approximation (SGA)). It is impossible to even mention all of them.

However, there is a problem with the canonical version of the $t-J$
model, since the projection onto the lowest Fock subspace (exclusion of the local double
occupancies $\{|i \uparrow \downarrow \rangle\}$)  invariably leads  to the non--fermion
anticommutation relation for the projected operators $\{ \hat{b}_{i\sigma}\}$ and $\{ \hat{b}_{j\sigma}^\dagger\}$. We proposed a formal resolution of this principal obstacle \cite{Spalek2017}. 

Namely, we can start from an extended $t-J$ model in the  $t-J-U$ form
\begin{align}\label{eq223}
    \hat{\widetilde{\mathcal{H}}} &= \sum_{ij\sigma}\!{'} t_{ij} \hat{a}_{i\sigma}^\dagger \hat{a}_{j\sigma} + \sum_{ij}\!{'}  J_{ij} (\hat{\mathbf{S}}_i \cdot \hat{\mathbf{S}}_j - \frac{1}{4} \hat{n}_{i}\hat{n}_{j}) \nonumber \\
    & + U \sum_i \hat{n}_{i\uparrow}\hat{n}_{i\downarrow}.
\end{align}

We see that we have added \textit{ad-hoc} the Hubbard term  to the original $t-J$ model. This trick allows us to omit the restricting projection onto the $P_1$ subspace, as we recover the full $t-J$ model in the strong correlation
limit when $U/|t| \gg 1$. However, the more inquisitive reader may ask if this formal trick 
is admissible from the physics perspective. The answer to this question is that it is correct within the general
perturbation expansion. Here, one formally takes $J_{ij}$ as an another parameter to 
be adjusted to the value $J_{ij} = 2 t_{ij}^2/U$ in the $|t|/U \ll 1$ limit. In any
case, in practically all papers concerning the theory of the high temperature superconductivity
$t_{ij}^2/U$ is fixed at the value $J_{\langle ij\rangle} /|t_{\langle ij \rangle}| \equiv J/|t|$ in the range 0.3 -- 0.4. Also, in such an approach one can take now a realistic value in the range of $U \sim 8 \div 12 ~eV$ for the high--temperature cuprates, not the much lower value obtained when one tries to map the corresponding many--band ($d-p$ model) onto its 
single--band version. This latter form of the $t-J-U$ model has been adopted in most of our recent
works.

\subsection{From local pairs to Cooper pairing and phase diagram}
The introduction  of the effective $t-J$ or $t-J-U$ Hamiltonian represents by itself 
only the first stage of the whole job. This is because we have introduced so far only a plausible 
effective Hamiltonian which provides an intuitive glance at what physical many--particle
(broken--symmetry) states may appear, but the detailed analysis requires separate solution methods. In other words, even though (\ref{eq223}) may be more attractive in that respect than the original Hubbard model, the determination of their ground
state or statistical--mechanical phases is equally, if not more complex than that of the first stage
developments.

A detailed discussion of the  solution is the subject of a series of papers \cite{Abram2017,Gutzwiller1965,Zegrodnik2017,Zegrodnik201795,Fidrysiak2018canonicalA,Zegrodnik2018a,Zegrodnik2018,Zegrodnik2019,Zegrodnik2020,Biborski2020,Zegrodnik2021}, as well as \cite{Spalek2022}, so we will not repeat it here. Instead, we delineate principal features of our approach and characterize differences with some of them, as well 
as present two specific results. 

We start from the Gutzwiller--type (SGA) approach to express the many--particle wave function, using a
projection operator proposed in the different form \cite{Spalek2017,Abram2017} than the original Gutzwiller one \cite{Gutzwiller1965}.
We then proceed next with the systematic expansion called diagrammatic 
expansion of the Gutzwiller (variational) approach (DE--GWF). Such a development 
constitutes our original approach to high--temperature
superconductivity \cite{Zegrodnik2017,Zegrodnik201795,Fidrysiak2018canonicalA,Zegrodnik2018a,Zegrodnik2018,Zegrodnik2019,Zegrodnik2020,Biborski2020,Zegrodnik2021,Fidrysiak2021c,Fidrysiak2021a,Fidrysiak2021}. The extended approach of DE-GWF \cite{Spalek2022,Spalek2017,Zegrodnik2017} results probably in the most involved 
phase diagram involving both d--wave superconductivity with a coexistent 
pair--density--wave order \cite{Spalek2022}, where also  a qualitative comparison
with the experimental data is displayed.

\begin{figure*}[!htbp]
\centering
\begin{tikzpicture}[
    x=0.95cm, y=0.95cm,
    font=\small,
    line cap=round,
    band edge/.style={line width=0.7pt},
    top edge/.style={line width=0.7pt,dash pattern=on 5pt off 3pt},
    level/.style={line width=0.65pt,dash pattern=on 4pt off 2.5pt},
    tick/.style={line width=0.7pt},
    extension/.style={line width=0.4pt},
    dimension/.style={
        line width=0.65pt,
        {Latex[length=2.0mm,width=1.35mm]}-{Latex[length=2.0mm,width=1.35mm]}
    },
    energy label/.style={anchor=east,inner sep=4pt},
    band label/.style={anchor=west,align=left,inner sep=0pt},
    d occupied/.style={pattern={Lines[
        angle=115,distance=4.3pt,line width=0.23pt]}},
    p occupied/.style={pattern={Lines[
        angle=45,distance=5.0pt,line width=0.23pt]}}
]
% ---------- PARAMETRY DO EDYCJI ----------
% Srodki i szerokosci energetyczne pasm:
\def\Ep{1.90}               % epsilon_p
\def\Ed{4.10}               % epsilon_d
\def\U{3.00}                % przesuniecie srodka gornego podpasma
\def\Wd{1.95}               % szerokosc kazdego podpasma d
\def\Wp{3.20}               % szerokosc pasma p
\def\Mu{4.86}               % potencjal chemiczny
% Szerokosci POZIOME sa wylacznie graficzne:
\def\XD{4.40}               % dlugosc poziomych krawedzi podpasm d
\def\XP{2.65}               % dlugosc poziomych krawedzi pasma p
\def\XDimD{4.77}            % polozenie strzalek W_d
\def\XDimP{2.90}            % polozenie strzalki W_p
\def\XText{6.00}            % polozenie nazw pasm
% ---------- GRANICE PASM ----------
\pgfmathsetmacro{\PLo}{\Ep-\Wp/2}
\pgfmathsetmacro{\PHi}{\Ep+\Wp/2}
\pgfmathsetmacro{\DLo}{\Ed-\Wd/2}
\pgfmathsetmacro{\DHi}{\Ed+\Wd/2}
\pgfmathsetmacro{\UpperCenter}{\Ed+\U}
\pgfmathsetmacro{\UpperLo}{\UpperCenter-\Wd/2}
\pgfmathsetmacro{\UpperHi}{\UpperCenter+\Wd/2}
\pgfmathsetmacro{\AxisTop}{\UpperHi+0.48}
\pgfmathsetmacro{\AxisBottom}{\PLo-0.55}
% Kreskowanie: obszary pasm ponizej mu.
\pgfmathsetmacro{\POcc}{max(\PLo,min(\Mu,\PHi))}
\pgfmathsetmacro{\DOcc}{max(\DLo,min(\Mu,\DHi))}
\pgfmathsetmacro{\UpperOcc}{max(\UpperLo,min(\Mu,\UpperHi))}

% ---------- KRESKOWANIE ----------
% W obszarze nalozenia pasm widoczne sa oba kierunki kreskowania.
\path[p occupied] (0,\PLo) rectangle (\XP,\POcc);
\path[d occupied] (0,\DLo) rectangle (\XD,\DOcc);
\path[d occupied] (0,\UpperLo) rectangle (\XD,\UpperOcc);

% ---------- KRAWEDZIE PASM ----------
\draw[top edge]  (0,\UpperHi) -- (\XD,\UpperHi);
\draw[band edge] (0,\UpperLo) -- (\XD,\UpperLo);
\draw[band edge] (0,\DHi) -- (\XD,\DHi);
\draw[band edge] (0,\DLo) -- (\XD,\DLo);
\draw[band edge] (0,\PHi) -- (\XP,\PHi);
\draw[band edge] (0,\PLo) -- (\XP,\PLo);

% ---------- OS ENERGII I POZIOMY ----------
\draw[line width=0.8pt,-{Latex[length=2.3mm,width=1.5mm]}]
    (0,\AxisBottom) -- (0,\AxisTop) node[above=2pt] {$E$};
\foreach \y/\lab in {
    \Ep/{\varepsilon_p},
    \Ed/{\varepsilon_d},
    \UpperCenter/{\varepsilon_d+U}}
{
    \draw[tick] (-0.095,\y) -- (0.11,\y);
    \node[energy label] at (-0.08,\y) {$\lab$};
}
\draw[level] (0,\Mu) -- (\XD,\Mu);
\node[energy label] at (-0.08,\Mu) {$\mu$};

% ---------- SZEROKOSCI ENERGETYCZNE ----------
\foreach \lo/\hi/\mid/\idx in {
    \UpperLo/\UpperHi/\UpperCenter/d,
    \DLo/\DHi/\Ed/d}
{
    \draw[extension] (\XD,\lo) -- ({\XDimD+0.10},\lo);
    \draw[extension] (\XD,\hi) -- ({\XDimD+0.10},\hi);
    \draw[dimension] (\XDimD,\lo) -- (\XDimD,\hi);
    \node[anchor=west,inner sep=4pt]
        at (\XDimD,\mid) {$W_{\idx}$};
}
\draw[extension] (\XP,\PLo) -- ({\XDimP+0.10},\PLo);
\draw[extension] (\XP,\PHi) -- ({\XDimP+0.10},\PHi);
\draw[dimension] (\XDimP,\PLo) -- (\XDimP,\PHi);
\node[anchor=west,inner sep=4pt] at (\XDimP,\Ep) {$W_p$};
\node[anchor=north west,inner sep=0pt,yshift=-5pt]
    at (\XP,\PLo) {$\varepsilon_p-W_p/2$};

% ---------- NAZWY PASM ----------
\node[band label] at (\XText,\UpperCenter)
    {upper Hubbard\\subband};
\node[band label] at (\XText,\Ed)
    {lower Hubbard\\subband};
\node[band label] at (\XText,\Ep)
    {$p$ band};
\end{tikzpicture}
\caption{Schematic model of the effective three d--p band structure after removing hybridization $d_{x^2-y^2}$ is less than half filled upon doping 
if p--band is completely filled even in the doped system. The condition
for the appearance of charge--transfer insulator are specified in main
text with including the p--p ($\sim U_p$) and p--d ($\sim U_{pd}$) interactions.
For explanation of such configuration of d band with respect to p band see the explanation
in main text, where p--p and p--d interactions are included in the discussion.}
\label{fig:6}
\end{figure*}
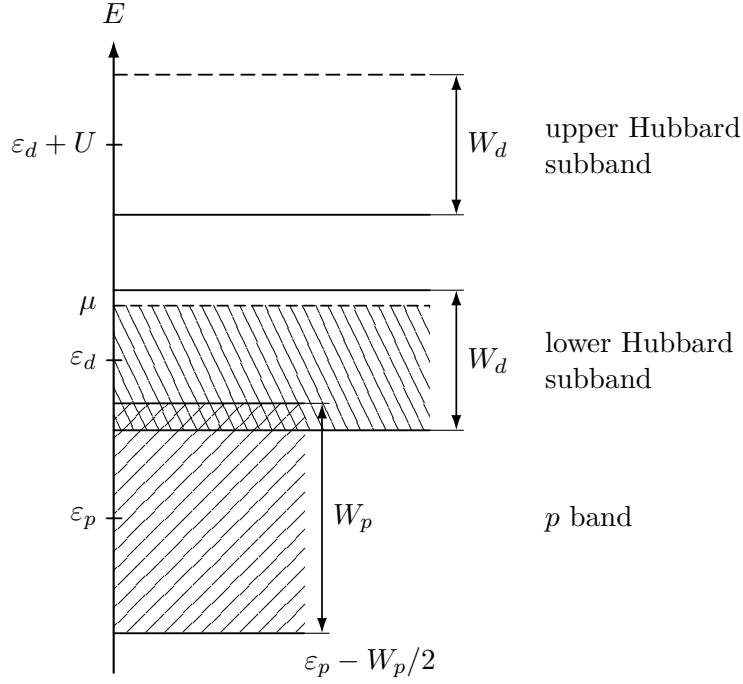
This solution within DE--GWF combined with extended $t-J-U$ model encompasses the following properties of high--temperature superconducting cuprates:
\begin{enumerate}[label=\roman* \!)]
\item A division into the BCS--like and non--BCS regimes of hole doping, the latter for the underdoped cuprates \cite{Spalek2017} (BCS stands for Bardeen, Cooper and Schrieffer).
\item The extended ($t-J-U$) model allows for a description of phase diagram on both hole-- ($n\leq1$)
and electron ($n\geq1$)--sides \cite{Zegrodnik2017}. 
\item Discussion of the role of interlayer single--particle tunneling on the phase diagram
within the full (DE--GWF) solution  \cite{Zegrodnik201795}. Also, realistic phase--diagram discussion \cite{Fidrysiak2018canonicalA}.
\item Appearance of a nematic phase within the extended model \cite{Zegrodnik2018a}.
\item Formulation of the reciprocal--space approach to DE--GWF, and discussion of the kink
in dispersion relation of electron excitations near the Fermi energy \cite{Fidrysiak2018canonicalA}.
\item The influence of charge--density--wave; and particularly pair--density--wave--excitation 
onto the phase diagram in the hole regime \cite{Zegrodnik2018}.
\item Diversion 1: Extensive study of the three--band d--p model and its relation to results obtained for single--band correspondent \cite{Zegrodnik2019,Zegrodnik2020,Biborski2020,Zegrodnik2021}.
\item Extension 1: Discussion of robust paramagnon and plasmon excitations within an
extended approach that includes Gaussian quantum fluctuations within SGA \cite{Fidrysiak2021c,Fidrysiak2021a,Fidrysiak2021}. Obviously, as with most papers on theoretical modeling, it is the experiment that determines which features of the model are correct. Such decisive test confirms the validity of the models, which as a rule  contain approximations. Also, the 
summary of most of the above results is provided in our review \cite{Spalek2022}.
\item Extension 2: Analysis of appearance of topological superconductivity within $t-J-U$ model
of twisted bilayer cuprates \cite{fidrysiak2023}.
\end{enumerate}

%%%%%%%%%%%%%%%%%%%%%%%%%%%%%%%%%%%%%%%%%%%%%%%%%%%%%%%%%%%%%%%%%%%%%%%%%%%%%%

\section{From three--band {{($d\!-\!p$)}} to single--band $t-J$ model?}
\subsection{The Emery model}
This question has been posed right at the beginning of research on superconductivity in the 
cuprates. It is also relevant in the context of heavy--fermion 
systems. Originally already in 1959 Anderson \cite{Anderson1959canonical} asked about the nature of the $3d$ electrons, since in the transition metal
oxides (CoO, NiO, MnO) the $3d$ ions with localized electrons are surrounded by the 
uncorrelated anions $O^{2-}$ with $2p$ electrons. The proposed answer was that the resultant $3d$
electrons are somewhat dressed with the $2p$--wave functions, but  their original atomic
space symmetry is preserved, e.g., $3d_{x^2-y^2}$, $3dz^2$, etc. The same concept is usually assumed 
in the situation with itinerant holes (empty) representing $3d$ states, when discussing one--band model.
\begin{figure}[htb]
\centering
\includegraphics[width=0.45\textwidth]{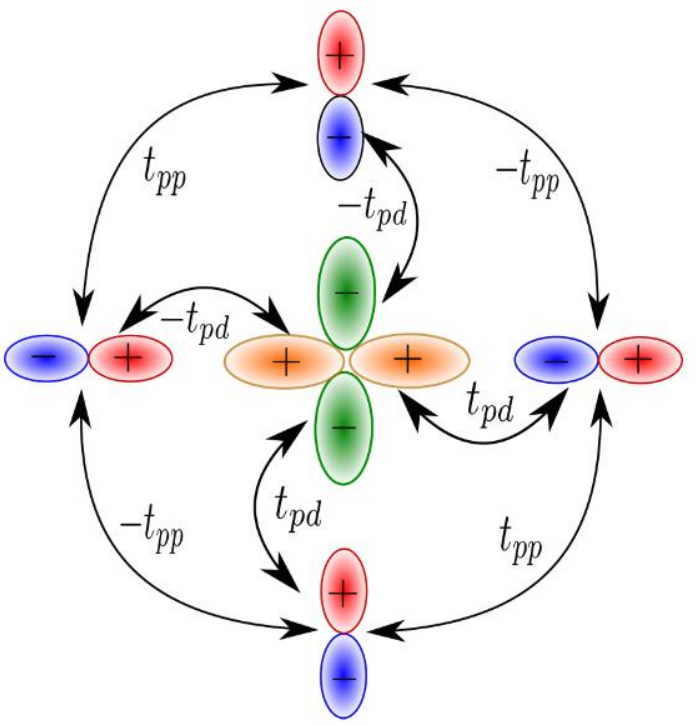}
\caption{The hopping parameters between the three types of orbitals in the model and the corresponding sign convention
for the antibonding orbital structure. The $d_{x^2-y^2}$ orbital is centered at the copper site and the $p_x/p_y$ orbitals 
are centered at the oxygen sites.}
\label{Fig4}
\end{figure}
This last presumption can be easily criticized as for the moving holes the Mott mechanism,
keeping $3d$ states, localized does not hold. Namely, a strong $3d-2p$ hybridization 
(quantum--mechanical mixing) is present and therefore one should expect that a three-band model
containing two $2p_{x,y}$ bands could be more appropriate. Such a model is schematically 
characterized in Fig. 4. This situation is commonly called the Emery model of the $CuO_2$ plane;
the Hamiltonian has the form \cite{Zegrodnik2019}
\begin{align}\label{eq31}
    \hat{\mathcal{H}} &= \sum_{il, jl',\sigma}\!{'} t_{ij}^{ll'} \hat{a}^{\dagger}_{il\sigma}\hat{a}_{il'\sigma} + \sum_{il} \epsilon_{il} (\epsilon_l - \mu)\hat{n}_{il} \nonumber \\
    &+ \sum_{il} U_l \hat{n}_{il\uparrow}\hat{n}_{il\downarrow} + \rm{interband ~ interactions},
\end{align}
where $\hat{a}^{\dagger}_{i l\sigma}$ ($\hat{a}_{i l\sigma}$) creates (annihilates) the electron with spin $\sigma$ at the $i-th$ atomic
site and orbital $l$ ($l \in 3d$,  then $\hat{a}^\dagger_{il\sigma} \equiv \hat{a}^\dagger_{i\sigma}$; $l \in p_x,~p_y$, then $\hat{a}^\dagger_{il\sigma} \equiv \hat{c}^\dagger_{i\sigma}$); the primed summation means that $i\neq j$, $l\neq l'$. The $2p$ orbitals in $Cu O_2$ plane are located at oxygen atomic sites which reside in between every two nearest-neighbors Cu sites containing $d_{x^2-y^2}$ orbital states (cf. Fig. 5). 
Also, for $l \in p_x,~p_y$, one has $ U_l \equiv U_p$, whereas for $l\in d_{x^2-y^2}~U_l\equiv U$. 

Additionally, for $l \neq l'$, $t^{l l'}_{\langle i j \rangle} \equiv t_{pd}$, whereas  for
$l \in p_x, ~ p_y$, one has $\epsilon_l \equiv \epsilon_p$ and for $l \in d_{x^2-y^2}$, $\epsilon_l = \epsilon_d$. Obviously, only can take instead $\epsilon_d - \epsilon_p$ and then $\hat{n}_{il}$ is replaced $\hat{n}_{id}$ and we can take $t^{ll}_{ij}$ as $t_{ij}^{pp}$ for $l\in p_x,~p_y$ and $t_{ij}^{dd} \simeq 0$. The single electron structure of the model
(\ref{eq31}) is shown in Fig. 4. One can then ask a principal question: How such a 3--band structure can be approximated by an effective single band--one? An obvious rationalization is that the $CuO_2^{2-}$ structural 
element contains a single $3d$ electron corresponding to the $Cu^{2+}$ ($3d^9)$) configuration,
if the eight $3d$ inner electrons can be regarded as inert  and the $2p$ states are coming from the filled $2p_x$ and $2 p_y$ shells. Out of this resultant five electrons, four fill the lower two
bands (bonding and nonbonding). Hence we are left with the remaining fifth electron which fills partially the uppermost antibonding band \cite{Zegrodnik2019}. Such interpretation justifies the sign 
convention of the hopping--integral configuration presented in Fig. 5. This view has been
rationalized by the similarity of results obtained using the one--band and three--band models, respectively \cite{Zegrodnik201795,Fidrysiak2018canonicalA,Zegrodnik2018a,Zegrodnik2018}.

\subsection{Atomic picture: Emergence of single $d$--band}
A different, but complementary view of the three--orbital electronic structure, resulting in an effective one d--band model, can be discussed as follows. The starting point here is the canonical transformation removing 
in the first order the hybridization term $\sim t_{pd}$ and replacing it by the effective interactions up to the fourth order. This is admissible
because in case of cuprates,  $|t_{pd}| \simeq 1.5 eV$ and is essentially smaller
than the $d-d$ Hubbard interaction $U_d \equiv U \simeq 8 \div 10 eV$. 
Also, carrying out the transformation starting from the Hamiltonian (3.1)
in real space, we avoid the  denominator singularities of the Schrieffer--Wolff--type of transformation \cite{Spalek1981a} since the distance between the $d$ and $p$ atomic levels is $|\epsilon_d - \epsilon_p| \simeq 3.5 eV > |t_{pd}|$ and $\epsilon_d - \epsilon_p + U \gg |t_{pd}|$. The additional
role between p and d electrons is played by the $p-p$ Coulomb interaction $U_{pp} \equiv U_p \sim 3 eV$ and
$U_{pd} \approx 1.5 eV$. Some of these values are taken from Ref. \cite{feiner1996}. The picture of the electronic 
structure is sketched in Fig. 4.

At first glance into Fig. 4 it may seem to look as if we have a Mott--Hubbard, not charge--transfer gap. However, when we include both $U_p$ and $U_{pd}$ terms, the
condition that charge gap appears that the p--d transition energy is 
lower than the direct $d-d$ transition between the Hubbard subbands. Taking
the situation in which one electron is taken either from the top of effective 
$p$ band or from the top of lower Hubbard subband, and transferred to the bottom of the upper Hubbard subband, we obtain the condition for the 
charge--transfer gap appearance, which reads
\begin{equation}
    \epsilon_d - \epsilon_p < \frac{W_p - W_d}{2} + U_p.
\end{equation}
This condition is easy to fulfill when $U_p \sim 3 eV$ for the parameters listed above the bare bandwidths $W_p$ and $W_d$ must be determined from
the explicit expression for the effective Hamiltonian \cite{dOrtoli2026}. 

However, there is still an additional requirement that the carriers (holes) are in the $d$--band, i.e., that $\epsilon_d + W/2 > \epsilon_p + W/2$; this 
leads to the condition 
\begin{equation}
    \epsilon_d - \epsilon_p > \frac{W_p + W_d}{2}.
\end{equation}

By comparing the two above inequalities we see that $U_p$ plays a crucial 
role in determining charge transfer character of the Mott--Hubbard insulator
and the $d$ character of the carriers.  Also, the effective $d$--level position is placed 
above that for $2p$ electrons, the point we discuss briefly below.

The above estimate should be put on a formal basis which 
explains, first of all, the origin of the $d$--band position and width $W_d$, as well as the nature of the interaction between d and p electrons at higher orders, 
namely the $d-p$ Kondo--type  and the $d-d$ kinetic exchange via $p$--orbitals
(superexchange). For that, we should rewrite Hamiltonian (3.1) in the more explicit form
\begin{align}\label{eq31}
    \hat{\mathcal{H}} &= \sum_{m,\sigma} \epsilon_p (\hat{c}^{\dagger}_{m 1 \sigma}\hat{c}_{m 1 \sigma} + \hat{c}^{\dagger}_{i 2 \sigma}\hat{c}_{i 2 \sigma}) \nonumber \\ &+ \sum_{i\sigma} \epsilon_{d} \hat{a}^{\dagger}_{i \sigma}\hat{a}_{i \sigma} 
    + \sum_{\langle i m \rangle} t_{pd} (\hat{c}^{\dagger}_{m 1 \sigma}\hat{a}_{i \sigma} + \hat{a}^{\dagger}_{i \sigma}\hat{c}_{m 1 \sigma}) \nonumber \\
    &+ \sum_{\langle i m \rangle \sigma} t_{pd} (\hat{c}^{\dagger}_{m 2 \sigma}\hat{a}_{i \sigma} + \hat{a}^{\dagger}_{i \sigma}\hat{c}_{m 2 \sigma})
    + U \sum_{i} \hat{n}_{i \uparrow}\hat{n}_{i \downarrow} \nonumber \\
    &+ U_p \sum_m (\hat{n}_{m 1 \uparrow}\hat{n}_{m 1 \downarrow} + \hat{n}_{m 2 \uparrow}\hat{n}_{m 2 \downarrow}),
\end{align}
where $\hat{c}$ operators with labels 1 or 2 represent original $2 p_x$ and 
$2 p_y$ orbitals, respectively.  Here we omit the details
of the canonical transformation that replaces the hybridization term $\!\equiv t_{pd}$ terms by the corresponding second--order Kondo interaction and the fourth--order superexchange \cite{dOrtoli2026}. Instead, for the purpose of the present 
discussion we only estimate them. Namely, the second--order terms providing 
the Kondo interaction between the pair $\langle i,m \rangle$ of sites is 
\begin{align}
    \sim & 2 t_{pd}^2 \left( \frac{1}{\epsilon_p - \epsilon_d - U_{p} + U_{pd}}
    + \frac{1}{\epsilon_p -\epsilon_d + U - U_p}\right) \nonumber \\
    &\times \sum_{\langle i m \rangle} (\hat{\mathbf{S}}_i \cdot \hat{\mathbf{s}}_m - \frac{1}{4} \hat{n}_i\hat{n}_m).
\end{align}
A similar form has the $d-d$ hopping via p state
\begin{align}
    \sim & 2 t_{pd}^2 \left( \frac{1 - n}{\epsilon_p - \epsilon_d - U_{p} + U_{pd}}
    + \frac{n}{\epsilon_p -\epsilon_d + U - U_p}\right) \nonumber \\
    & \times \sum_{\langle ij \rangle \sigma}  \hat{a}^{\dagger}_{i \sigma}\hat{a}_{j \sigma}.
\end{align}
So, the $d-d$ hopping ($d_i \to p_m \to d_{j\neq i}$) and the Kondo interaction
are of comparable magnitude. This basic conclusion clarifies the concept of the Zhang--Rice (Kondo) singlet contribution to lowering the position of the $d$--level. Namely, the concept of the Kondo--singlet local binding competes
with the $d$--band formation which likely may destroy the singlet.
At the  same time, we introduce the nonzero $d$--electron 
bandwidth roughly $\sim (1 - n)$. The fourth--order $d-d$ superexchange, as well as detailed 
analysis of the role of the Kondo interaction is deferred to a separate publication \cite{dOrtoli2026}.

\section{A brief summary}
Any modeling of the quasi--two--dimensional high--$T_c$ cuprates must respect a
few basic experimental facts. First of them is that a strongly correlated metallic
state evolves from a Mott insulating state upon doping. This state
is destroyed upon a small doping ($\delta \simeq 0.05$), not by any Kondo compensating cloud of $p$ electrons. The last raised here is the fact that the Mott insulator is of the charge--transfer type, even though the carriers in
the doped state are presumably of $d$ character. It must be so since simultaneously assume that the carriers are strongly correlated and interact via kinetic exchange interaction. The last feature constitutes the essence of the $t-J$ model.

All the above requirements are met by the situation sketched schematically in Fig. 4, where we obtain an effective $t-J$ model with an additional Kondo--type coupling, and itineracy of $d$ electrons. However, the Kondo--type coupling vanishes if all electrons are accommodated in the $d$--band, i.e., if $\langle \hat{n}_i \rangle \simeq n_d$, $\langle \hat{n}_m \rangle \simeq n_c = 0$. Otherwise, the Kondo--type part can be
transformed into the hybrid--pairing form which should be also of $d$--wave type \cite{Spalek1988a}. But then, strictly
speaking, the three--orbital model is more appropriate. This point should be clarified further as the latter situation (with $n_c \neq 0$) may lead to the situation in which with the Zhang--Rice singlet is directly related to the narrow $d$--band dynamics. We should see a clarification
along this line soon, so that the role of the one--band $t-J$ model in the context of high--$T_c$ 
superconductivity can be (hopefully) finally resolved.
This is particularly important, since the $t-J$($-U$) model provides a good overall description of the 
principal properties of high--$T_c$ cuprates, including the dynamical (paramagnon and plasmon) excitations \cite{Spalek2022,Fidrysiak2021c,Fidrysiak2021a,Fidrysiak2021}.
%\cite{Spalek2026A1}
%\section{Discussion}

%\section{Conclusions}

\section*{Acknowledgments}
The work was supported by the Narodowe Centrum Nauki (NCN) Grant No. 2023/49/B/ST3/03545.
The author would like to thank Andrzej M. Oleś and Koung-an Chao, who helped in finishing
a many--month struggle to get published the original work on $t-J$ model.
Thanks are also due to Dr. Maciek Fidrysiak for discussion and Dr. Piotrek Kuterba, with additional thanks to the latter for technical help.
\clearpage
\appendix
\section{$t-J$ model in general form}
In the strong correlation limit the $t-J$ model was originally derived starting from the assumption that the Fock space of the relevant states is decomposed into two subspaces, namely the lower containing only the empty or singly occupied sites (Wannier states) and the upper, containing the doubly occupied states. This concept reflects qualitatively the concept of 
the Hubbard subbands. As the splitting between those states is sizable and of the order of
U -- W, where $W=z|\sum_{j(i)} t_{ij}|$ is the bare bandwidth, we can regard the hopping involving the doubly occupied states as highly excited. We can remove these states (\ref{eq19}) by 
canonical perturbation expansion  and replace them by virtual hopping processes back and
forth in higher order. In the second order this procedure leads to the effective Hamiltonian.
Under the above conditions the complete (extended) $t-J$ Hamiltonian reads
\begin{strip}
\begin{align} \label{eq21}
    P_1 \widetilde{H} P_1 & \quad = P_1 \Bigr\{ \sum_{i j \sigma} t_{ij} \hat{a}^\dagger_{i \sigma} (1 - \hat{n}_{i \overline{\sigma}}) \hat{a}_{j \sigma} (1 - \hat{n}_{j \overline{\sigma}})  \nonumber \\
    & \quad + \sum_{ij} \frac{2 t^2_{ij}}{U} \left[ \hat{\mathbf{S}}_i \cdot \hat{\mathbf{S}}_j - \frac{1}{4} \sum_{\sigma \sigma'} \hat{n}_{i \sigma} (1 - \hat{n}_{i \overline{\sigma}}) \hat{n}_{j \sigma} (1 - \hat{n}_{j \overline{\sigma}}) \right]   \nonumber \\
    & \quad + \sum_{ijk \sigma \sigma'} \frac{t_{ij} t_{jk}}{ U} \hat{a}^\dagger_{j \sigma'} (1 - \hat{n}_{j \overline{\sigma}'}) \hat{a}_{k\sigma'} \hat{n}_{k \overline{\sigma}} \hat{a}^\dagger_{k \sigma} \hat{n}_{k\overline{\sigma}} \hat{a}_{i\sigma} (1 - \hat{n}_{i \overline{\sigma}})\Bigr\} P_1.
\end{align}
\end{strip}
\printbibliography

@Book{Dirac2010canonical,
  author    = {Dirac, P. A. M.},
  publisher = {Clarendon Press, Oxford University Press},
  title     = {The principles of quantum mechanics},
  year      = {2010},
  address   = {Oxford },
  edition   = {4. ed. (rev.), repr.},
  isbn      = {9780198520115},
  %note      = {Hier auch später erschienene, unveränderte Nachdrucke},
  number    = {27},
  series    = {International series of monographs on physics},
  pagetotal = {314},
  ppn_gvk   = {1602991790},
}

@Misc{Spalek2020sorbitals,
  note = {For $s$--orbitals, those microscopic parameters have been
          evaluated rigorously within the EDABI method, e.g.,
          J.~Spa{\l}ek, ``Mott Physics in Correlated Nanosystems:
          Localization--Delocalization Transition by the Exact
          Diagonalization Ab Initio Method,'' in
          \textit{Topology, Entanglement, and Strong Correlations},
          E.~Pavarini and E.~Koch (Eds.),
          Forschungszentrum J{\"u}lich Zentralbibliothek, Verlag,
          J{\"u}lich 2020, pp.~7.1--7.38.}
}

@Article{Spalek1988,
  author    = {Spałek, J.},
  journal   = {Phys. Rev. B},
  title     = {Effect of pair hopping and magnitude of intra-atomic interaction on exchange-mediated superconductivity},
  year      = {1988},
  issn      = {0163-1829},
  month     = jan,
  number    = {1},
  pages     = {533--536},
  volume    = {37},
  doi       = {10.1103/physrevb.37.533},
  publisher = {American Physical Society (APS)},
}

@Article{Zegrodnik2017,
  author    = {Zegrodnik, Michał and Spałek, Józef},
  journal   = {Phys. Rev. B},
  title     = {{Universal properties of high-temperature superconductors from real-space pairing: Role of correlated hopping and intersite Coulomb interaction within the t-J-U model}},
  year      = {2017},
  issn      = {2469-9969},
  month     = aug, 
  number    = {5},
  pages     = {054511},
  volume    = {96},
  doi       = {10.1103/physrevb.96.054511},
  publisher = {American Physical Society (APS)},
}

@Article{Dey2026a,
author    = {Dey, Tushar and Fidrysiak, Maciej and Spałek, Józef},
  journal   = {},
  title     = {},
  year      = {2026},
  issn      = {},
  number    = {},
  pages     = {},
  volume    = {},
  doi       = {},
  publisher = {},
  note    = { to be published}
}

@Article{Zegrodnik2018,
  author    = {Zegrodnik, Michał and Spałek, Józef},
  journal   = {Physical Review B},
  title     = {Incorporation of charge- and pair-density-wave states into the one-band model of d -wave superconductivity},
  year      = {2018},
  issn      = {2469-9969},
  month     = oct,
  number    = {15},
  pages     = {155144},
  volume    = {98},
  doi       = {10.1103/physrevb.98.155144},
  publisher = {American Physical Society (APS)},
}

@Article{Fidrysiak2018canonicalA,
  author    = {Fidrysiak, M. and Zegrodnik, M. and Spałek, J.},
  journal   = {J. of Phys.: Condens. Matter},
  %year      = {2018},
  month     = may,
  %pages     = {475602},
  volume    = {30},
  note = {475602 (2018), and Refs. therein}
}

@Article{Fidrysiak2021,
  author    = {Fidrysiak, Maciej and Spałek, Józef},
  journal   = {Phys. Rev. B},
  title     = {Unified theory of spin and charge excitations in high- Tc cuprate superconductors: A quantitative comparison with experiment and interpretation},
  year      = {2021},
  issn      = {2469-9969},
  month     = jul,
  number    = {2},
  pages     = {l020510},
  volume    = {104},
  doi       = {10.1103/physrevb.104.l020510},
  publisher = {American Physical Society (APS)},
}

@Article{Fidrysiak2021a,
  author    = {Fidrysiak, Maciej and Spałek, Józef},
  journal   = {Phys. Rev. B},
  title     = {{Universal collective modes from strong electronic correlations: Modified 1/N theory with application to high- Tc cuprates}},
  year      = {2021},
  issn      = {2469-9969},
  month     = apr,
  number    = {16},
  pages     = {165111},
  volume    = {103},
  doi       = {10.1103/physrevb.103.165111},
  publisher = {American Physical Society (APS)},
}

@Article{Spalek1988a,
  author    = {Spałek, Józef},
  journal   = {Phys. Rev. B},
  title     = {{Microscopic model of hybrid pairing: A common approach to heavy-fermion and high-Tc superconductivity}},
  year      = {1988},
  issn      = {0163-1829},
  month     = jul,
  number    = {1},
  pages     = {208--212},
  volume    = {38},
  doi       = {10.1103/physrevb.38.208},
  publisher = {American Physical Society (APS)},
}

@Article{Zegrodnik2019,
  author    = {Zegrodnik, M. and Biborski, A. and Fidrysiak, M. and Spałek, J.},
  journal   = {Phys. Rev. B},
  title     = {Superconductivity in the three-band model of cuprates: Variational wave function study and relation to the single-band case},
  year      = {2019},
  issn      = {2469-9969},
  month     = mar,
  number    = {10},
  pages     = {104511},
  volume    = {99},
  doi       = {10.1103/physrevb.99.104511},
  publisher = {American Physical Society (APS)},
}

@Article{Spalek2010,
  author    = {Spałek, J. and Kurzyk, J. and Podsiadły, R. and Wójcik, W.},
  journal   = {Eur. Phys. J. B},
  title     = {{Extended Hubbard model with the renormalized Wannier wave functions in the correlated state II: quantum critical scaling of the wave function near the Mott-Hubbard transition}},
  year      = {2010},
  issn      = {1434-6036},
  month     = mar,
  number    = {1},
  pages     = {63--74},
  volume    = {74},
  doi       = {10.1140/epjb/e2010-00077-6},
  publisher = {Springer Science and Business Media LLC},
}

@Article{Spalek2022,
  author    = {Spałek, J. and Fidrysiak, M. and Zegrodnik, M. and Biborski, A.},
  journal   = {Phys. Rep.},
  title     = {{Superconductivity in high-Tc and related strongly correlated systems from variational perspective: Beyond mean field theory}},
  year      = {2022},
  issn      = {0370-1573},
  month     = may,
  pages     = {1--117},
  volume    = {959},
  doi       = {10.1016/j.physrep.2022.02.003},
  publisher = {Elsevier BV},
}

@Article{Chao1977canonical,
  author    = {Chao, K A and Spałek, J and Oleś, A M},
  journal   = {Journal of Physics C: Solid State Physics},
  title     = {Kinetic exchange interaction in a narrow S-band},
  year      = {1977},
  issn      = {0022-3719},
  month     = may,
  number    = {10},
  pages     = {L271--L276},
  volume    = {10},
  doi       = {10.1088/0022-3719/10/10/002},
  publisher = {IOP Publishing},
}

@article{spalek1977magnetic,
  title={Magnetic phases of tight-binding electrons including intraatomic exchange},
  author={Spa{\l}ek, J. and Ole{\'s}, A. M.},
  journal={Physica B+C},
  volume={86--88},
  pages={375--377},
  year={1977},
  publisher={Elsevier}
}

@Article{Spalek2017,
  author    = {Spałek, Józef and Zegrodnik, Michał and Kaczmarczyk, Jan},
  journal   = {Phys. Rev. B},
  title     = {Universal properties of high-temperature superconductors from real-space pairing: {t-J-U} model and its quantitative comparison with experiment},
  year      = {2017},
  issn      = {2469-9969},
  month     = jan,
  number    = {2},
  pages     = {024506},
  volume    = {95},
  doi       = {10.1103/physrevb.95.024506},
  publisher = {American Physical Society (APS)},
}

@Article{Anderson1959canonical,
  author    = {Anderson, P. W.},
  journal   = {Physical Review},
  title     = {New Approach to the Theory of Superexchange Interactions},
  year      = {1959},
  issn      = {0031-899X},
  month     = jul,
  number    = {1},
  pages     = {2--13},
  volume    = {115},
  doi       = {10.1103/physrev.115.2},
  publisher = {American Physical Society (APS)},
}

@InBook{Anderson1963,
  author    = {Anderson, Philip W.},
  pages     = {99--214},
  publisher = {Elsevier},
  title     = {Theory of Magnetic Exchange Interactions:Exchange in Insulators and Semiconductors},
  year      = {1963},
  volume = {14},
  editor = {Seitz, F. and Turnbull, S.},
  isbn      = {9780126077148},
  booktitle = {Solid State Physics},
  doi       = {10.1016/s0081-1947(08)60260-x},
  issn      = {0081-1947},
}

@Article{Hubbard1964canonical,
  author    = {Hubbard, J.},
  journal   = {Proceedings of the Royal Society of London. Series A. Mathematical and Physical Sciences},
  title     = {{Electron correlations in narrow energy bands III. An improved solution}},
  year      = {1964},
  issn      = {2053-9169},
  month     = sep,
  number    = {1386},
  pages     = {401--419},
  volume    = {281},
  doi       = {10.1098/rspa.1964.0190},
  publisher = {The Royal Society},
}

@Article{Gutzwiller1965,
  author    = {Gutzwiller, Martin C.},
  journal   = {Phys. Rev.},
  title     = {Correlation of Electrons in a Narrow Band},
  year      = {1965},
  issn      = {0031-899X},
  month     = mar,
  number    = {6A},
  pages     = {A1726--A1735},
  volume    = {137},
  doi       = {10.1103/physrev.137.a1726},
  publisher = {American Physical Society (APS)},
}

@Article{Hubbard1963,
  author    = {Hubbard, J.},
  journal   = {Proceedings of the Royal Society of London. Series A. Mathematical and Physical Sciences},
  title     = {Electron correlations in narrow energy bands},
  year      = {1963},
  issn      = {2053-9169},
  month     = nov,
  number    = {1365},
  pages     = {238--257},
  volume    = {276},
  doi       = {10.1098/rspa.1963.0204},
  publisher = {The Royal Society},
}

@Article{Hubbard1965,
  author    = {Hubbard, J.},
  journal   = {Proceedings of the Royal Society of London. Series A. Mathematical and Physical Sciences},
  title     = {Electron correlations in narrow energy bands {IV}. The atomic representation},
  year      = {1965},
  issn      = {0080-4630},
  month     = may,
  number    = {1403},
  pages     = {542--560},
  volume    = {285},
  doi       = {10.1098/rspa.1965.0124},
}

@Book{Abrikosov1963,
  author     = {Abrikosov, A. A. and Gor'kov, L. P. and Dzyaloshinski, I. E.},
  publisher  = {Prentice-Hall},
  title      = {Methods of Quantum Field Theory in Statistical Physics},
  year       = {1963},
  address    = {Englewood Cliffs, NJ},
  isbn       = {9780135785591}%,
  %translator = {Silverman, Richard A.},
}

@Book{PinesNozieres1966,
  author    = {Pines, David and Nozi{\`e}res, Philippe},
  publisher = {W. A. Benjamin},
  title     = {{The Theory of Quantum Liquids. Volume I: Normal Fermi Liquids}},
  year      = {1966},
  address   = {New York},
  %pagetotal = {355},
}

@Article{Luttinger1960,
  author  = {Luttinger, J. M.},
  journal = {Physical Review},
  title   = {Fermi Surface and Some Simple Equilibrium Properties of a System of Interacting Fermions},
  year    = {1960},
  month   = aug,
  number  = {4},
  pages   = {1153--1163},
  volume  = {119},
  doi     = {10.1103/PhysRev.119.1153},
}

@Article{Wigner1934,
  author  = {Wigner, E.},
  journal = {Physical Review},
  title   = {On the Interaction of Electrons in Metals},
  year    = {1934},
  month   = dec,
  number  = {11},
  pages   = {1002--1011},
  volume  = {46},
  doi     = {10.1103/PhysRev.46.1002},
}

@Book{FockSelectedWorks2004,
  author    = {Fock, V. A.},
  editor    = {Faddeev, L. D. and Khalfin, L. A. and Komarov, I. V.},
  publisher = {Chapman \& Hall/CRC},
  title     = { Selected Works: Quantum Mechanics and Quantum Field Theory},
  year      = {2004},
  address   = {Boca Raton, FL},
  edition   = {1},
  isbn      = {9780415300025},
  pagetotal = {584},
  doi       = {10.1201/9780203643204},
}

@Article{Heisenberg1928,
  author  = {Heisenberg, W.},
  journal = {Zeitschrift f{\"u}r Physik},
  title   = {Zur Theorie des Ferromagnetismus},
  year    = {1928},
  month   = sep,
  number  = {9--10},
  pages   = {619--636},
  volume  = {49},
  doi     = {10.1007/BF01328601},
}

@Article{SpalekOles1976,
  author      = {Spa{\l}ek, J{\'o}zef and Ole{\'s}, Andrzej M.},
  %institution = {Jagiellonian University},
  title       = {{On the kinetic exchange interactions in the Hubbard model}},
  %year        = {1976},
  %a%ddress     = {Krak{\'o}w},
  %month       = oct,
  number      = {SSPJU 6/76},
  %pages       = {1--12},
  %type        = {Preprint},
  %doi         = {10.26106/4s8w-hv63},
  note        = {Jagiellonian University, Kraków, Oct. 1976, pp. 1–12; doi:10.26106/4s8w-hv63},
}

@Article{Fock1932,
  author   = {Fock, V.},
  journal  = {Zeitschrift für Physik},
  title    = {Konfigurationsraum und zweite Quantelung},
  year     = {1932},
  volume   = {75},
  pages    = {622--647},
  doi      = {10.1007/BF01344458},
  addendum = {Reprinted in English as
              \emph{Configuration Space and Second Quantization},
              in \emph{V. A. Fock: Selected Works:
              Quantum Mechanics and Quantum Field Theory},
              edited by L. D. Faddeev, L. A. Khalfin,
              and I. V. Komarov
              (Chapman \& Hall/CRC, Boca Raton, 2004),
              pp.~191--220},
}

@Article{SpalekChaoOles1978,
  author  = {Spa{\l}ek, J. and Chao, K. A. and Ole{\'s}, A. M.},
  journal = {Physics Letters A},
  title   = {Antiferromagnetism of strongly correlated electrons in narrow bands},
  year    = {1978},
  month   = jun,
  number  = {6},
  pages   = {503--506},
  volume  = {66},
  doi     = {10.1016/0375-9601(78)90411-5},
}

@Misc{Spalek1981Habilitation,
  author       = {Spa{\l}ek, J{\'o}zef},
  title        = {{Habilitation Thesis}},
  year         = {1981},
  howpublished = {Jagiellonian University, Krak{\'o}w},
  note         = {Unpublished},
}

@Article{Spalek1978a,
  author    = {Spałek, J. and Oleś, A. M. and Chao, K. A.},
  journal   = {physica status solidi (b)},
  title     = {The Effective Magnetic Interactions between Impurity and Conduction Electrons for the Wolff Model},
  year      = {1978},
  issn      = {1521-3951},
  month     = June,
  number    = {2},
  pages     = {625--635},
  volume    = {87},
  doi       = {10.1002/pssb.2220870228},
  publisher = {Wiley},
}

@Misc{Spalek2026SCQMoverview,
  note = {For a brief overview see, e.g.,
          J.~Spa{\l}ek, ``Strongly Correlated Quantum Matter:
          {$t$--$J$} Model, Real-Space Pairing,
          Spin-Dependent Masses, and Atomicity in Chemical Bond
          and Nanosystems,'' \textit{Acta Physica Polonica B}
          \textbf{57}, 5-A1 (2026),
          doi:10.5506/APhysPolB.57.5-A1.}
}

@article{Anderson1987,
author = {P. W. Anderson },
title = {The Resonating Valence Bond State in $La_2CuO_4$ and Superconductivity},
journal = {Science},
volume = {235},
number = {4793},
pages = {1196-1198},
year = {1987},
doi = {10.1126/science.235.4793.1196},
URL = {https://www.science.org/doi/abs/10.1126/science.235.4793.1196},
eprint = {https://www.science.org/doi/pdf/10.1126/science.235.4793.1196}}

@article{SpalekGoc2012,
doi = {10.1088/0031-8949/86/04/048301},
url = {https://doi.org/10.1088/0031-8949/86/04/048301},
year = {2012},
month = {sep},
publisher = {IOP Publishing},
volume = {86},
number = {4},
pages = {048301},
author = {Spałek, Jozef and Goc-Jagło, Danuta},
title = {On the strongly correlated quantum matter paradigm: magnetism–superconductivity redux},
journal = {Physica Scripta}
}

@article{Abram2017,
doi = {10.1088/1361-648X/aa7a21},
url = {https://doi.org/10.1088/1361-648X/aa7a21},
year = {2017},
month = {aug},
publisher = {IOP Publishing},
volume = {29},
number = {36},
pages = {365602},
author = {Abram, M and Zegrodnik, M and Spałek, J},
title = {Antiferromagnetism, charge density wave, and d-wave superconductivity in the extended {t-J-U} model: role of intersite Coulomb interaction and a critical overview of renormalized mean field theory},
journal = {Journal of Physics: Condensed Matter}
}

@article{Zegrodnik201795,
  title = {{Effect of interlayer processes on the superconducting state within the t-J-U model: Full Gutzwiller wave-function solution and relation to experiment}},
  author = {Zegrodnik, Micha\l{} and Spa\l{}ek, J\'ozef},
  journal = {Phys. Rev. B},
  volume = {95},
  issue = {2},
  pages = {024507},
  numpages = {9},
  year = {2017},
  month = {Jan},
  publisher = {American Physical Society},
  doi = {10.1103/PhysRevB.95.024507},
  url = {https://link.aps.org/doi/10.1103/PhysRevB.95.024507}
}

@article{Zegrodnik2018a,
doi = {10.1088/1367-2630/aac6f7},
url = {https://doi.org/10.1088/1367-2630/aac6f7},
year = {2018},
month = {jun},
publisher = {IOP Publishing},
volume = {20},
number = {6},
pages = {063015},
author = {Zegrodnik, Michał and Spałek, Józef},
title = {Stability of the coexistent superconducting-nematic phase under the presence of intersite interactions},
journal = {New Journal of Physics}
}

@Article{Zegrodnik2020,
  author  = {Zegrodnik, Micha{\l} and Biborski, Andrzej and Spa{\l}ek, J{\'o}zef},
  title   = {Superconductivity and intra-unit-cell electronic nematic phase
             in the three-band model of cuprates},
  journal = {Eur. Phys. J. B},
  volume  = {93},
  pages   = {183},
  year    = {2020},
  doi     = {10.1140/epjb/e2020-10290-3}
}

@article{Biborski2020,
  title = {Superconducting properties of the hole-doped three-band $d\ensuremath{-}p$ model studied with minimal-size real-space $d$-wave pairing operators},
  author = {Biborski, A. and Zegrodnik, M. and Spa\l{}ek, J.},
  journal = {Phys. Rev. B},
  volume = {101},
  issue = {21},
  pages = {214504},
  numpages = {9},
  year = {2020},
  month = {Jun},
  publisher = {American Physical Society},
  doi = {10.1103/PhysRevB.101.214504},
  url = {https://link.aps.org/doi/10.1103/PhysRevB.101.214504}
}

@article{Zegrodnik2021,
doi = {10.1088/1361-648X/abcff6},
url = {https://doi.org/10.1088/1361-648X/abcff6},
year = {2021},
month = {aug},
publisher = {IOP Publishing},
volume = {33},
number = {41},
pages = {415601},
author = {Zegrodnik, M and Biborski, A and Fidrysiak, M and Spałek, J},
title = {Superconductivity in the three-band model of cuprates: nodal direction characteristics and influence of intersite interactions},
journal = {Journal of Physics: Condensed Matter}
}

@Article{Fidrysiak2021c,
  author    = {Fidrysiak, Maciej and Goc-Jagło, Danuta and Spałek, Józef},
  journal   = {Journal of Magnetism and Magnetic Materials},
  title     = {{Collective spin and charge excitations in the t-J-U model of high-Tc cuprates}},
  year      = {2021},
  issn      = {0304-8853},
  month     = Dec,
  pages     = {168395},
  volume    = {539},
  doi       = {10.1016/j.jmmm.2021.168395},
  publisher = {Elsevier BV},
}

@article{fidrysiak2023,
  title = {{Tuning topological superconductivity within the t-J-U model of twisted bilayer cuprates}},
  author = {Fidrysiak, Maciej and Rzeszotarski, Bart\l{}omiej and Spa\l{}ek, J\'ozef},
  journal = {Phys. Rev. B},
  volume = {108},
  issue = {22},
  pages = {224509},
  numpages = {13},
  year = {2023},
  month = {Dec},
  publisher = {American Physical Society},
  doi = {10.1103/PhysRevB.108.224509},
  url = {https://link.aps.org/doi/10.1103/PhysRevB.108.224509}
}

@article{Spalek1981a,
title = {Split-band antiferromagnet with the intermediate valence},
journal = {Solid State Communications},
volume = {37},
number = {7},
pages = {571-574},
year = {1981},
issn = {0038-1098},
doi = {https://doi.org/10.1016/0038-1098(81)90136-8},
url = {https://www.sciencedirect.com/science/article/pii/0038109881901368},
author = {J. Spałek and A.M. Oleś}
}

@article{feiner1996,
  title = {{Effective single-band models for the high-${\mathit{T}}_{\mathit{c}}$ cuprates. I. Coulomb interactions}},
  author = {Feiner, L. F. and Jefferson, J. H. and Raimondi, R.},
  journal = {Phys. Rev. B},
  volume = {53},
  issue = {13},
  pages = {8751--8773},
  numpages = {0},
  year = {1996},
  month = {Apr},
  publisher = {American Physical Society},
  doi = {10.1103/PhysRevB.53.8751},
  url = {https://link.aps.org/doi/10.1103/PhysRevB.53.8751}
}

@Misc{dOrtoli2026,
  author = {d'Ortoli Galerneau, F. and Spa{\l}ek, J{\'o}zef},
  note   = {in preparation}
}

@Misc{JedrakSpalek2010,
  note = {J.~J{\k e}drak, J.~Spa{\l}ek,
          \textit{Phys. Rev. B} \textbf{81}, 073108 (2010).},
}

@Misc{JedrakSpalek2011,
  note = {J.~J{\k e}drak, J.~Spa{\l}ek,
          \textit{Phys. Rev. B} \textbf{83}, 104512 (2011).},
}
\end{document}